\documentclass[aps,pra,twocolumn,showpacs,superscriptaddress, nofootinbib]{revtex4-2}

\usepackage[english]{babel}
\usepackage[utf8]{inputenc}

\usepackage[pdftex, pdftitle={Article}, pdfauthor={Author}]{hyperref} 

\usepackage{natbib}
\usepackage{hyperref}
\usepackage{verbatim}
\usepackage{amsmath}
\usepackage{amsfonts}
\usepackage{amssymb}
\usepackage{amsthm}
\usepackage{mathtools}
\usepackage{bm}
\usepackage{xparse}
\usepackage{graphicx}
\usepackage{xcolor}
\usepackage{textcase}
\usepackage{url}
\usepackage{lipsum}
\usepackage{appendix}
\usepackage{epstopdf}

\def\Tr{\mathrm{Tr}}

\newcommand{\diagdots}[3][-25]{%
  \rotatebox{#1}{\makebox[0pt]{\makebox[#2]{\xleaders\hbox{$\cdot$\hskip#3}\hfill\kern0pt}}}%
}

\DeclareDocumentCommand{\Tr}{m m O{\big}}{{\rm Tr}_{\:\!{#1}}#3({#2}#3)}

\begin{document}
\title{Comment on  ``Quantum principle of relativity"}

\author{Flavio Del Santo}
\affiliation{Vienna Center for Quantum Science and Technology (VCQ), Faculty of Physics, Boltzmanngasse 5, University of Vienna, Vienna A-1090, Austria.}
\affiliation{Institute for Quantum Optics and Quantum Information (IQOQI),
Austrian Academy of Sciences, Boltzmanngasse 3, A-1090 Vienna, Austria.}

\author{Sebastian Horvat}
\thanks{The authors contributed equally.}
\affiliation{Vienna Center for Quantum Science and Technology (VCQ), Faculty of Physics, Boltzmanngasse 5, University of Vienna, Vienna A-1090, Austria.}
\date{\today}

\begin{abstract}
A. Dragan and A. Ekert [\textit{New Journal of Physics}, 22(3), p.033038.] have recently claimed that fundamental properties of quantum physics (e.g. fundamental indeterminism and the principle of superposition) can be derived solely from relativistic considerations, if one takes as physically meaningful superluminal reference frames. In this comment we show that their arguments  are flawed and their claims therefore unwarranted.

\end{abstract}
\maketitle

\section{Introduction}

In a recent paper, A. Dragan and A. Ekert \cite{dragan2020quantum} advanced the hypothesis that some of the signature features of quantum theory can be derived solely from special relativity, by considering a general derivation of the Lorentz transformations which includes superluminal terms that are usually discarded on physical grounds. 
In particular, their main claim, broadly construed, is that taking superluminal reference frames seriously would lead to (i) fundamental indeterminism, (ii) the superposition principle and --with a few more mathematical assumptions-- (iii) the complex probability amplitudes, all key characteristics of quantum physics.

As recalled in Ref.\cite{dragan2020quantum},  the Galilean principle of relativity allows to derive the two following sets of relativistic transformations (in 1+1 dimensions), both of which preserve the constancy of the speed of light $c$: the standard Lorentz transformations
\begin{equation}
\begin{aligned}
x'= \frac{x-Vt}{\sqrt{1-V^2/c^2}} \\
t'= \frac{t-Vx/c^2}{\sqrt{1-V^2/c^2}},
\end{aligned}
\end{equation}\\
which are well behaved for $V<c$, and a second class of transformations
\begin{equation}
\label{trans}
\begin{aligned}
x'= \pm \frac{V}{|V|} \frac{x-Vt}{\sqrt{V^2/c^2-1}} \\
t'=  \pm \frac{V}{|V|} \frac{t-Vx/c^2}{\sqrt{V^2/c^2-1}},
\end{aligned}
\end{equation}\\
that holds instead for velocities $V>c$.
Dragan and Ekert's work aims to investigate the physical consequences of taking as physically meaningful superluminal reference frames and therefore the transformations given by Eq. \eqref{trans}, allegedly showing that these lead to the aforementioned (quantum) properties.

However, in what follows we show that even if one is willing to give serious consideration to faster-than-light reference frames (and we do not see anything wrong per se with exploring this possibility), the arguments that purport to derive the above features from the principle of relativity --we will limit our analysis to the first two points-- are either untenable or require additional assumptions, which would be in need of further independent justification. We therefore conclude that the claims made in Ref. \cite{dragan2020quantum} are unwarranted.

\section{The arguments and our counterarguments}
\subsection{Fundamental indeterminism}
\begin{figure}
\includegraphics[width=\linewidth]{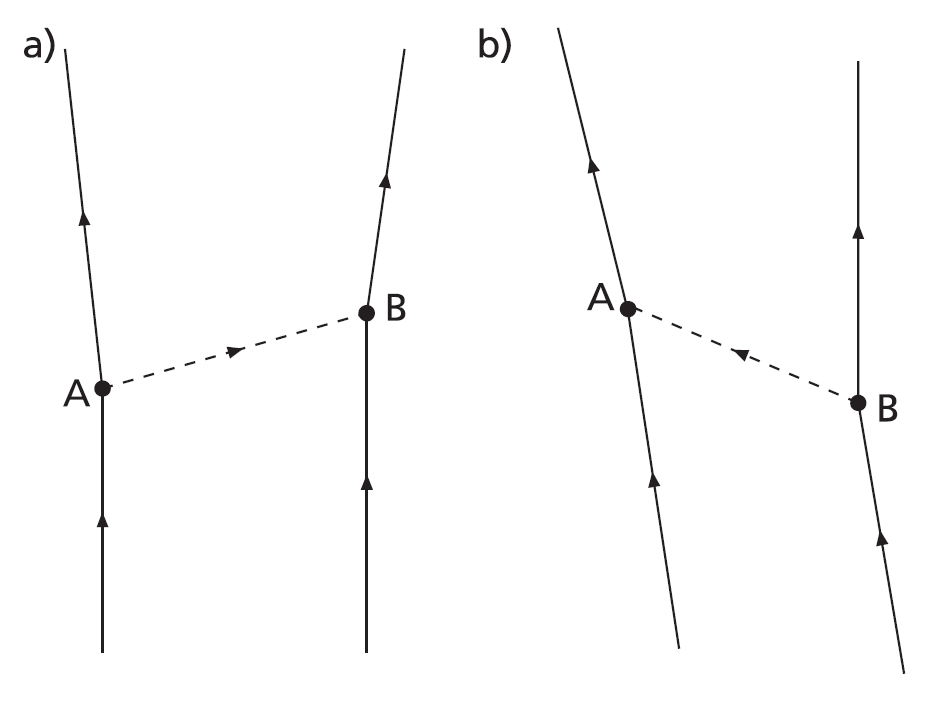}
\caption{``Spacetime diagrams of a process of sending a superluminal particle as seen by two inertial observers (time is vertical, space is horizontal): (a) particle emitted from A and absorbed in B, (b) the same process observed in a different inertial frame.'' Figure and caption taken from Ref. \cite{dragan2020quantum}.}
\centering
\label{fig1}
\end{figure}

\begin{figure}
\includegraphics[width=\linewidth]{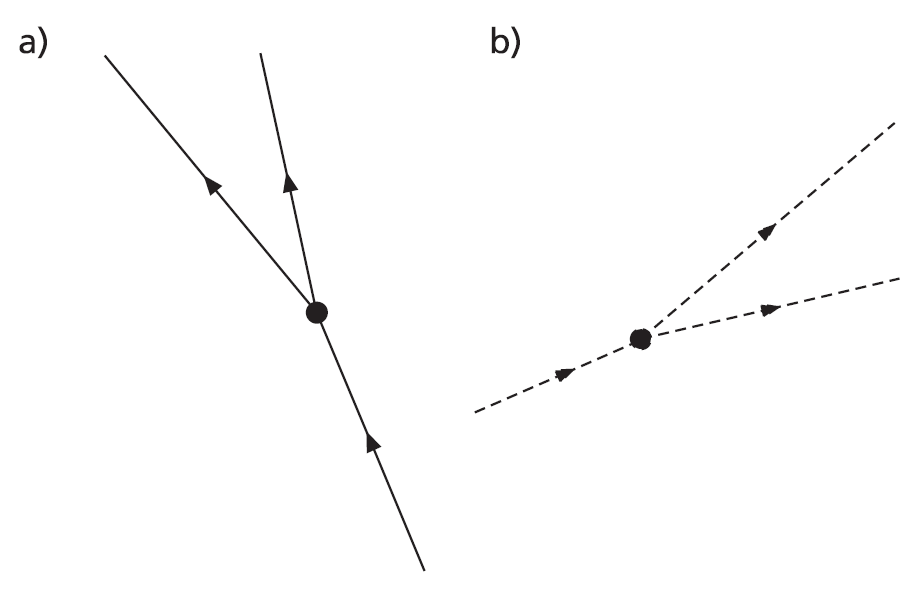}
\caption{``A spacetime diagram of a decay of a subluminal particle into a pair of subluminal particles (time is vertical, space is horizontal): (a) in a subluminal reference frame, (b) in a superluminal reference frame.'' Figure and caption taken from Ref. \cite{dragan2020quantum}.}
\centering
\label{fig2}
\end{figure}

The first argument put forward in Ref. \cite{dragan2020quantum} is that of indeterminism. The authors consider a superluminal particle (tachyon) being emitted at the spacetime point A and absorbed later at a spacelike separated point B, as described within a given inertial reference frame $O$. In another reference frame $O'$, related via a subluminal velocity to $O$, the particle is described as being emitted at point B and absorbed at A. The two descriptions are represented as spacetime diagrams in Fig. \ref{fig1}. 
Let us focus on the frame $O$ and assume that the moment of emission of the particle is fully determined by the properties of the events contained within its past light-cone, i.e. that there exists a local and deterministic model of the particle emission. On the other hand, switching to frame $O'$, it seems that the emission of the particle at B is not determined by any events lying in its corresponding past light-cone: the emission of the particle is thus ``completely spontaneous and fundamentally unpredictable", thereby rendering a local and deterministic account impossible. The authors finally claim that the only way to preserve the Galilean principle of relativity --and thus avoid introducing a preferred reference frame-- is to conclude that no ``local and deterministic description of the emission of a superluminal particle is possible in any inertial frame" \cite{dragan2020quantum}.
Moreover, since the transformations \eqref{trans} allow to turn a superluminal signal into a subluminal one and vice-versa --as seen by observers in a reference frame that moves with relative superluminal velocity-- the argument for fundamental indeterminism should apply not only to tachyons but more generally: e.g. there can be no local and deterministic model of a subluminal particle decaying into two subluminal particles, as pictured in Fig. \ref{fig2}.\\
The above argument can be reconstructed as the following entailment of conclusions C1 and C2 by premises P1-P3:\\
\textbf{P1}: The Galilean principle of relativity implies that, if a local deterministic model of a physical process cannot be given in one inertial reference frame, then it cannot be given in any other inertial reference frame.\\
\textbf{P2}: A physical process consisting in a superluminal particle being emitted and absorbed at two spacelike separated events (as in Fig. \ref{fig1}) \textit{seems} to allow a local deterministic model in reference frame $O$, but definitely does not allow a local and deterministic model in reference frame $O'$.\\
\textbf{P3}: Any particle can be regarded to travel superluminally relative to a class of inertial reference frames.\\
\textbf{C1}: By P1 and P2, the emission of superluminal particles does not allow a local deterministic model and is thus fundamentally indeterministic.\\
\textbf{C2:} By C1 and P3, the decay of subluminal particles does not allow a local deterministic model and is thus fundamentally indeterministic as well.\footnote{Notice that we added the qualifier \textit{seems} in premise P2, as the argument would otherwise result in a logical contradiction, i.e. ``there \textit{is} a local deterministic model within reference frame $O$, and there \textit{is no} local deterministic model within the reference frame $O$''.}\\
In what follows, we will show that the above argument contains several problems.\\
\textbf{Remark.} First of all, we want to remark that premise P1 requires further justification and specification. The Galilean principle of relativity, as usually construed, states that the equations of motion need to retain the same form in all inertial reference frames, and is thus not explicitly related to the (non)existence of local deterministic models in various reference frames. The authors should thus state precisely what they mean by ``the Galilean principle of relativity'' and provide a reason for why it is to be related to the (non)existence of local deterministic models. While giving such a definition is surely viable, we think there might be problems with justifying it due to the following reason. Notice that `local' in `local deterministic' refers explicitly to the standard relativistic definition of locality: an event is explained locally and deterministically if its occurrence is fully determined by the events lying in its past light-cone. However, we think that there is no physical reason to give preference to events lying in the past light-cone to those lying outside of it, since we are considering both subluminal and superluminal reference frames on the same footing. Thus, even though the notion of ``local determinism'' is surely well defined in this context, there is no reason for it to figure in the definition (or to be a consequence) of the Galilean principle of relativity.\\
\textbf{Counterargument 1.} Furthermore, granted that one accepts P1, one ought to also accept the following premise: ``If a local deterministic model of a physical process \textit{can} be given in one inertial reference frame, then it must be possible to provide one in all other inertial reference frames.'' Denying the latter would imply an a priori preference for non-local-deterministic models, whereas a principle of relativity (such as the Galilean one) should only have a claim on relational features between different frames. One should thus replace P1 with: ``Either all inertial reference frames allow local deterministic models, or none of them do''.

With the latter in mind, premise P2 turns out suspicious as well: at pains of avoiding a logical contradiction, we needed to add that it merely \textit{seems} that there is a local deterministic model in frame $O$, but that it \textit{is definitely the case} that there is none in frame $O'$. However, one may turn the story around and claim that it is definitely the case that there \textit{is} a local deterministic model in frame $O$ and that it merely \textit{seems} as if there was none in frame $O'$. Again, there is no positive reason to accept the former rather than the latter. A staunch proponent of local determinism may indeed argue analogously to the authors that, since per assumption there exists a local deterministic model in frame $O$, the Galilean principle of relativity guarantees that there is one in frame $O'$ as well (despite the deceiving appearance that there is none, i.e. it may be a model of \textit{hidden variables}). Such an argument would again be erroneous, unless its proponent provided an explicit local deterministic model in reference frame $O$. All in all, conclusions C1 and C2 are not warranted due to P1 being incomplete (it prefers non-local-deterministic models) and P2 being unjustified (the non-existence of a local deterministic model in frame $O'$ is not argued for).\\
\textbf{Counterargument 2.} Even if one were to ignore the above problems and to fully accept the reasoning provided by the authors, the argument leads to untenable consequences. The authors have clearly intended for the argument to apply only to microscopic phenomena (e.g. particle decays); however, the Galilean principle of relativity, as standardly construed, can be applied to macroscopic bodies as well. Indeed, there is no positive reason to prevent one from considering the diagrams in Figs. \ref{fig1} and \ref{fig2} as representations of macroscopic bodies (e.g. billiard balls, human beings or galaxies) moving and colliding in spacetime. For example, instead of a particle decay, Fig. \ref{fig2} may represent a rock containing a dynamite set to explode exactly 1 minute after being prepared, thereby splitting the rock into two parts: the authors' argument would lead one to conclude that the detonation of the rock was fundamentally indeterministic. However, that is obviously not the case, as not only that physicists can provide faithful local and deterministic models of the dynamite, but engineers can even use these models to construct the dynamite itself. The point is that the conclusions provided by the authors are intended to apply exclusively to those phenomena that are adequately modelled by quantum theory; however, this domain does not coincide with the domain of phenomena to which the Galilean principle of relativity is applicable. This exhibits a fundamental deficiency in any attempt of deriving features of quantum theory \textit{exclusively} from relativistic principles.\\
\textbf{Counterargument 3.} The authors argue for fundamental indeterminism by attempting to display an inconsistency between local determinism and Galilean relativity (when extended to superluminal reference frames). However, the argument can be reiterated by replacing ``local deterministic models'' with ``local probabilistic models'' (or simply ``local models''), provided that one accepts the following modification of premise P1: ``Either all inertial reference frames allow local probabilistic models, or none of them do''. Indeed, there is no reason to accept the relation between local deterministic models and Galilean relativity without also accepting the relation between the latter and local probabilistic models.\footnote{As explained in the previous counterarguments, we would deny both connections.} According to the authors, from the perspective of reference frame $O'$, ``the past world-line of the particle B carries no information about the time of the event B'', implying that ``the emission at B was completely spontaneous and fundamentally unpredictable'' \cite{dragan2020quantum}. This would then imply that also within frame $O'$, the emission at A must have been completely spontaneous and fundamentally unpredictable. On the other hand, if one were able to provide a local probabilistic model in one of the reference frames, then one would be able to fix the probability distribution of the emission event, based on the events lying in its past light-cone, and this distribution may as well sometimes be peaked around a certain value (indeed, recall that, mathematically speaking, deterministic models can be regarded as a subclass of probabilistic models, i.e. those that assign probabilities 0 or 1 to all events). Therefore, the argument put forward by the authors does not lead merely to the negation of local determinism, but to the negation of any quantitative local explanation whatsoever (be it deterministic or probabilistic). The issue of (in)determinism is thus irrelevant for this discussion: only locality (as standardly conceived in relativity) is at stake when considering superluminal reference frames, which is not that surprising.

Incidentally, by putting forward their argument for indeterminism, the authors also seem to be trying to give a more solid physical ground to their proposal of considering superluminal velocities. In fact, they maintain that ``if we had a source of superluminal particles at our disposal, we would not be able to use it to send any information because we would not be able to control the emission rate using any local operations" \cite{dragan2020quantum}. However, this claim appears to be justified only in the case that the indeterminacy of the source is indeed \textit{maximal}, i.e., when there is no possibility even in principle to have any information whatsoever on \textit{when} or even \textit{if} there will be an emission. On the other hand, indeterministic theories that deserve any credit are usually provided with some measure of likelihood. For instance, quantum mechanics, which is considered the indeterministic theory \textit{par excellence},\footnote{There are, however, further proposals of fundamentally indeterministic theories, including the alternative indeterministic interpretations of classical physics \cite{del2019physics, del2021indeterminism} and of special relativity \cite{del2021relativity}, proposed by one of us.} allows to compute the probability of events, such us the decay of a radioactive nucleus (to whose emitted particles we grant here the possibility in principle to travel faster than light), which can then be used to signal. It would suffice to agree between two parties that the time window for detecting a signal is long enough to allow a decay with a probability close to certainty. 

\subsection{The superposition principle}

The second main claim of Dragan and Ekert is that from the principle of relativity, one can retrieve another property characteristic of quantum theory, namely ``the fact that a particle that is not observed behaves as if it was moving along multiple trajectories at once'' \cite{dragan2020quantum} (superposition principle).

\begin{figure}
\includegraphics[width=\linewidth]{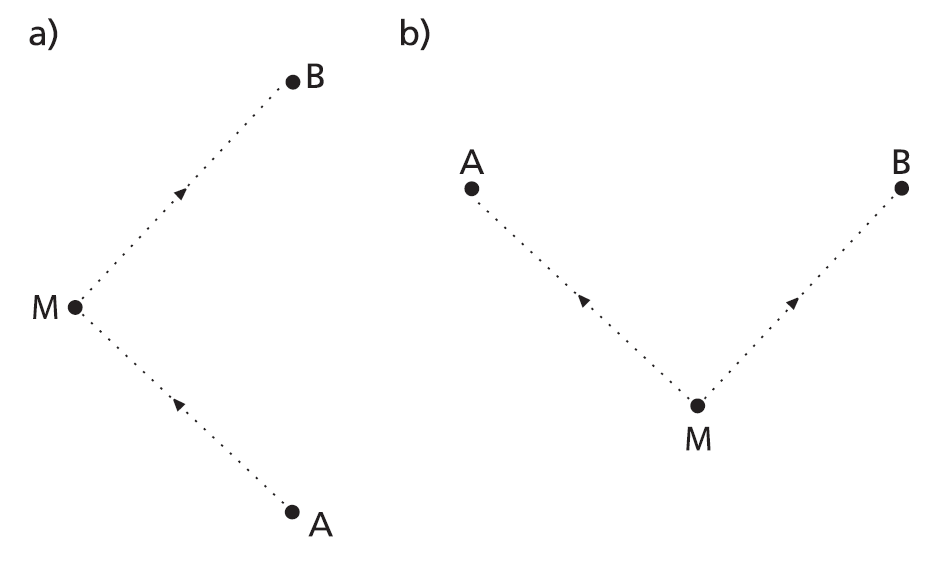}
\caption{``A spacetime diagram of a luminal particle (dotted line) reflected from a mirror (time is vertical, space is horizontal): (a) in a subluminal reference frame, (b) in a superluminal reference frame.'' Figure and caption taken from Ref. \cite{dragan2020quantum}.}
\centering
\label{fig3}
\end{figure}

To support this, they propose the following simple arrangement. In the rest frame of an observer $O$, consider a physical object that travels at the speed of light from a source (event $A$), bounces on a mirror $M$ and is reflected back to the original position (event $B$). This object has a well-defined single trajectory, as shown in the spacetime diagram in Fig. \ref{fig3}.  
Dragan and Ekert proceed by noticing that for an infinite velocity, the transformations \eqref{trans} reduce to    
\begin{equation}
\label{transinf}
\begin{aligned}
x'= ct \\
ct'= x. 
\end{aligned}
\end{equation}\\
Thus an observer $O'$ moving at infinite speed with respect to the reference frame of $O$, will see the time and the space axes swapped. In this new frame, the same set-up will look as if the object originated at $M$ and ``followed two trajectories'' towards the (now spatially separate) locations $A$ and $B$. The authors then state: ``If a detector placed at the path A-M absorbs the photon [=the object in question], then a similar detector placed at the path M-B will not register anything, because the photon has been absorbed earlier. Similarly, if a detector at M-B absorbed the photon, then certainly, the photon could not have been detected at the path A-M.'' \cite{dragan2020quantum}. Therefore, despite associating two paths M-A and M-B to the object in question, observer $O'$ still regards the process to involve a single object (or particle), since, if she were to introduce two absorbing detectors in the set-up, then only one of them would click (i.e. register the object), thereby resembling a quantum particle in spatial superposition. From the latter they conclude that ``even if we start with an idea of a classical particle moving along a single path, it is only a matter of a change of the reference frame to arrive at a scenario involving more than one path'' \cite{dragan2020quantum}.\\
The argument may be formulated as follows:\\
\textbf{P1}: While a classical object (e.g. particle) may follow a well defined spacetime trajectory in one reference frame, there exist other inertial frames, namely those related to the first one via superluminal velocities, relative to which multiple trajectories are associated to the same object.\\
\textbf{P2}: When one places absorbing detectors along the path of the object, one observes anti-correlations between the detection events.\\
\textbf{C}: Therefore, a classical object moving along a single trajectory in one reference frame can behave as a quantum particle in spatial superposition in another reference frame.\\
We think that this argument is flawed due to the following reasons.\\
\textbf{Counterargument 1.} Let us start with a criticism of P2 and its relation to C. Namely, the authors invoke explicitly only absorbing detectors, i.e. those that destroy the object in question upon detecting it. The reason they do so is to introduce an anti-correlation between the detection events. This is however a very stringent assumption, as there are surely detectors that do not destroy their pertaining objects upon registering them. Therefore, the argument seemingly applies exclusively to those objects which are necessarily destroyed upon measurement (i.e. those objects for which a non-demolishing measurement \textit{does not exist}). 
This is why the authors of Ref. \cite{dragan2020quantum}, provide as the only example that of a photon, which arguably is always destroyed upon measurement. This, however, jeopardises the strength of their argument, for a photon is a genuinely quantum concept while the aim of the authors is to retrieve a property of quantum physics starting only from classical relativistic considerations.

At any rate, even if the argument were to be applied to objects necessarily destroyed upon measurement (like perhaps photons) only, one may simply construct a device that absorbs and re-emits the object in question (or creates a new one). The argument thus does not hold even for photons, let alone for neutrons or electrons, which are commonly modelled by quantum mechanics, and for which non-demolishing measurements do exist.\\
\textbf{Counterargument 2.} There is a further fundamental problem. Consider a standard quantum particle (e.g. electron) prepared in an equally weighted spatial superposition of ``two paths'', and two demolishing detectors placed at the end of their corresponding paths. Elementary quantum mechanics and experimental practice tell us that each detector registers the particle with probability $\frac{1}{2}$, and that if one detector clicks, the other does not. Now consider again the scenario proposed by Dragan and Ekert in Fig. \ref{fig3}, and suppose that one places one detector in the middle of path A-M and another in path M-B. It trivially follows from the left picture that the detector placed in A-M will click with probability 1, and the second detector will never click (because the particle is absorbed at the previous detector). In the superluminal reference frame, the following thus holds: (i) if both detectors are present, only the one in path M-A clicks; (ii) if only one detector is present, then it clicks with probability 1. Therefore, the set-up considered by the authors does not replicate at the empirical level the behaviour of a quantum particle in spatial superposition.\\
\textbf{Counterargument 3.} The authors' argument relies on the fact that a single trajectory is mapped into two trajectories via a superluminal reference frame transformation. There are two problems with the latter reasoning.\\
Firstly, the diagram in Fig. \ref{fig3}b) does not need to be interpreted as representing two trajectories M-A and M-B; one may alternatively understand it as still representing a single trajectory that consists of the particle travelling backwards in time from A to M, and then forwards in time from M to B. A proponent for the latter interpretation may even provide positive reasons for such a view: (i) it preserves the causal order between the events, as the particle travels from A towards B with an intermediate bounce at M in all reference frames, and (ii) M still behaves as a mirror in all reference frames. On the other hand, concerning point (ii), Dragan and Ekert's view arguably implies that we can turn a mirror into a particle source (or into a beam-splitter) merely by changing our reference frame, which, while not logically inconsistent, leaves one wanting for further explanation. Anyhow, our aim is not to endorse the above alternative interpretation, but just to emphasize that more interpretations are possible and that the authors ought to argue why their view is the most preferable one (without invoking any quantum mechanical principles, as that would just beg the question).

The second problem is far more serious: there is a whole class of quantum superposition phenomena that cannot be captured by a mere change of reference frame. For example, consider a standard Mach-Zehnder experiment that consists in a particle travelling in a ``superposition of 
two paths'', both of which meet at the same spacetime location (at a beamsplitter), thereby closing a spacetime loop. In order to see that such an experiment cannot be accounted for by the authors' argument, it is enough to notice that a reference frame transformation (be it subluminal or superluminal) cannot turn an open trajectory into a closed one. The same holds for any other phenomenon involving loops. In the paper, the authors actually mention that such trajectories can be obtained, but do not provide any detailed explanation for it. A further problem is that a single trajectory traced by a classical particle can be transformed only to those trajectories which can be bijectively mapped into a real interval (e.g. the interval $[0,1]$). Therefore, one could not even account for a particle entering into a beamsplitter and ``splitting into two paths'', which would trace a trajectory in the shape of the letter `Y'. This is also the reason that mirror M in Fig. \ref{fig3}b) spontaneously produces a particle, without any object entering into it from the past. \\
\textbf{Counterargument 4.} A generic comment analogous to Counterargument 2 from the previous section can be made here as well. The authors intend to use special relativity to infer that microscopic objects follow the quantum superposition principle; however, relativistic arguments may be applied to a much wider domain of objects. For instance, one may take Fig. \ref{fig3}a) to represent a billiard ball bouncing from the edge of a billiard table, which would be described within a superluminal reference frame as the billiard ball ``travelling in superposition''; however, billiard balls manifestly travel through definite trajectories. One may arguably model even a billiard ball as a quantum mechanical system; however, the ball would then be represented as interacting with the table on which it is rolling and with the remaining environment, which would induce decoherence and highly suppress non-classical effects. 

\section{Conclusion}
In this comment, we have shown that Dragan and Ekert's claims in Ref. \cite{dragan2020quantum} are unwarranted. We have provided detailed counterarguments against their allegation that certain quantum features --fundamental indeterminism and the principle of superposition-- can be derived only by invoking the Galilean principle of  relativity and taking seriously the superluminal class of relativistic transformations that follow from the latter.
We did not deem necessary to discuss here the further claim in \cite{dragan2020quantum}, according to which, the complex probability amplitudes which characterize quantum mechanics can be inferred solely from relativistic considerations, because this is quite logically independent from the line of thought commented in the previous sections, and does not rely on the generalised transformations \eqref{trans}.

In conclusion, while it is surely desirable to try to derive the peculiar features of quantum theory from some fundamental principles, it does not seem that the principles of special relativity alone are sufficient for this program.

\section*{Acknowledgements} \label{sec:acknowledgements}
We thank \v Caslav Brukner and Borivoje Daki\' c for useful comments. F.D.S. acknowledges the financial support  from
European Innovation Council and SMEs Executive Agency,
TEQ 766900 / Project OEUP0259. S.H. acknowledges financial support from the Austrian
Science Fund (FWF) through BeyondC-F7112.

\newpage

\bibliography{Refs}

\begin{thebibliography}{4}%
\makeatletter
\providecommand \@ifxundefined [1]{%
 \@ifx{#1\undefined}
}%
\providecommand \@ifnum [1]{%
 \ifnum #1\expandafter \@firstoftwo
 \else \expandafter \@secondoftwo
 \fi
}%
\providecommand \@ifx [1]{%
 \ifx #1\expandafter \@firstoftwo
 \else \expandafter \@secondoftwo
 \fi
}%
\providecommand \natexlab [1]{#1}%
\providecommand \enquote  [1]{``#1''}%
\providecommand \bibnamefont  [1]{#1}%
\providecommand \bibfnamefont [1]{#1}%
\providecommand \citenamefont [1]{#1}%
\providecommand \href@noop [0]{\@secondoftwo}%
\providecommand \href [0]{\begingroup \@sanitize@url \@href}%
\providecommand \@href[1]{\@@startlink{#1}\@@href}%
\providecommand \@@href[1]{\endgroup#1\@@endlink}%
\providecommand \@sanitize@url [0]{\catcode `\\12\catcode `\$12\catcode
  `\&12\catcode `\#12\catcode `\^12\catcode `\_12\catcode `\%12\relax}%
\providecommand \@@startlink[1]{}%
\providecommand \@@endlink[0]{}%
\providecommand \url  [0]{\begingroup\@sanitize@url \@url }%
\providecommand \@url [1]{\endgroup\@href {#1}{\urlprefix }}%
\providecommand \urlprefix  [0]{URL }%
\providecommand \Eprint [0]{\href }%
\providecommand \doibase [0]{https://doi.org/}%
\providecommand \selectlanguage [0]{\@gobble}%
\providecommand \bibinfo  [0]{\@secondoftwo}%
\providecommand \bibfield  [0]{\@secondoftwo}%
\providecommand \translation [1]{[#1]}%
\providecommand \BibitemOpen [0]{}%
\providecommand \bibitemStop [0]{}%
\providecommand \bibitemNoStop [0]{.\EOS\space}%
\providecommand \EOS [0]{\spacefactor3000\relax}%
\providecommand \BibitemShut  [1]{\csname bibitem#1\endcsname}%
\let\auto@bib@innerbib\@empty
\bibitem [{\citenamefont {Dragan}\ and\ \citenamefont
  {Ekert}(2020)}]{dragan2020quantum}%
  \BibitemOpen
  \bibfield  {author} {\bibinfo {author} {\bibfnamefont {A.}~\bibnamefont
  {Dragan}}\ and\ \bibinfo {author} {\bibfnamefont {A.}~\bibnamefont {Ekert}},\
  }\bibfield  {title} {\bibinfo {title} {Quantum principle of relativity},\
  }\href@noop {} {\bibfield  {journal} {\bibinfo  {journal} {New Journal of
  Physics}\ }\textbf {\bibinfo {volume} {22}},\ \bibinfo {pages} {033038}
  (\bibinfo {year} {2020})}\BibitemShut {NoStop}%
\bibitem [{\citenamefont {Del~Santo}\ and\ \citenamefont
  {Gisin}(2019)}]{del2019physics}%
  \BibitemOpen
  \bibfield  {author} {\bibinfo {author} {\bibfnamefont {F.}~\bibnamefont
  {Del~Santo}}\ and\ \bibinfo {author} {\bibfnamefont {N.}~\bibnamefont
  {Gisin}},\ }\bibfield  {title} {\bibinfo {title} {Physics without
  determinism: Alternative interpretations of classical physics},\ }\href@noop
  {} {\bibfield  {journal} {\bibinfo  {journal} {Physical Review A}\ }\textbf
  {\bibinfo {volume} {100}},\ \bibinfo {pages} {062107} (\bibinfo {year}
  {2019})}\BibitemShut {NoStop}%
\bibitem [{\citenamefont {Del~Santo}(2021)}]{del2021indeterminism}%
  \BibitemOpen
  \bibfield  {author} {\bibinfo {author} {\bibfnamefont {F.}~\bibnamefont
  {Del~Santo}},\ }\bibfield  {title} {\bibinfo {title} {Indeterminism,
  causality and information: Has physics ever been deterministic?},\ }in\
  \href@noop {} {\emph {\bibinfo {booktitle} {Undecidability, Uncomputability,
  and Unpredictability}}}\ (\bibinfo  {publisher} {Springer},\ \bibinfo {year}
  {2021})\ pp.\ \bibinfo {pages} {63--79}\BibitemShut {NoStop}%
\bibitem [{\citenamefont {Del~Santo}\ and\ \citenamefont
  {Gisin}(2021)}]{del2021relativity}%
  \BibitemOpen
  \bibfield  {author} {\bibinfo {author} {\bibfnamefont {F.}~\bibnamefont
  {Del~Santo}}\ and\ \bibinfo {author} {\bibfnamefont {N.}~\bibnamefont
  {Gisin}},\ }\bibfield  {title} {\bibinfo {title} {The relativity of
  indeterminacy},\ }\href@noop {} {\bibfield  {journal} {\bibinfo  {journal}
  {Entropy}\ }\textbf {\bibinfo {volume} {23}},\ \bibinfo {pages} {1326}
  (\bibinfo {year} {2021})}\BibitemShut {NoStop}%
\end{thebibliography}%

\end{document}